\documentclass[a4paper,11pt]{article}
\pdfoutput=1 
\usepackage{jcappub} 
\usepackage{gensymb}
\usepackage[]{algorithm2e}
\usepackage{subcaption}
\usepackage{sidecap}
\usepackage{aasmacros}
\usepackage{todonotes}

\usepackage[T1]{fontenc} 
\newcommand{\kll}{\left(}
\newcommand{\klr}{\right)}

\newcommand{\fieldsize}{3.81^{\circ} \times 3.81^{\circ}}
\newcommand{\sskew}{\ensuremath{\mathcal{S}_{\kappa}}}
\newcommand{\kkurt}{\ensuremath{\mathcal{K}_{\kappa}}}

\newcommand{\loss}{{\ell}}
\newcommand{\Loss}{{\mathcal L}}
\newcommand{\w}{{\bf w}}
\newcommand{\W}{{\mathbf W}}
\newcommand{\x}{{\bf x}}
\newcommand{\ybar}{\bar{y}}

\notoc

\usepackage{array}
\newcolumntype{L}[1]{>{\raggedright\let\newline\\\arraybackslash\hspace{0pt}}m{#1}}
\newcolumntype{C}[1]{>{\centering\let\newline\\\arraybackslash\hspace{0pt}}m{#1}}
\newcolumntype{R}[1]{>{\raggedleft\let\newline\\\arraybackslash\hspace{0pt}}m{#1}}
\title{\boldmath Cosmological model discrimination with Deep Learning}

 \author[a]{J. J. K. Schmelzle,}
 \author[b]{A. Lucchi,}
 \author[a]{T. Kacprzak,}
 \author[a]{A. Amara,}
 \author[a]{R. Sgier,}
 \author[a]{A. R\'{e}fr\'{e}gier,}
 \author[b]{and T. Hofmann}
 \affiliation[a]{Department of Physics, Eidgen\"ossische Technische Hochschule Z\"urich,\\Wolfgang-Pauli-Str. 27, 8093 Z\"{u}rich, Switzerland}
  \affiliation[b]{Data Analytics Lab, Eidgen\"ossische Technische Hochschule Z\"urich,\\Universit\"atstrasse 6, 8092 Z\"{u}rich, Switzerland}

\emailAdd{sjorit@ethz.ch}
\emailAdd{aurelien.lucchi@inf.ethz.ch}
\emailAdd{tomasz.kacprzak@phys.ethz.ch}
\emailAdd{adam.amara@phys.ethz.ch}
\emailAdd{rsgier@phys.ethz.ch}
\emailAdd{alexandre.refregier@phys.ethz.ch}
\emailAdd{thomas.hofmann@inf.ethz.ch}

\keywords{gravitational lensing - dark matter - deep learning - cosmological model discrimination}

\abstract{
We demonstrate the potential of Deep Learning methods for measurements of cosmological parameters from density fields, focusing on the extraction of non-Gaussian information.
We consider weak lensing mass maps as our dataset.
We aim for our method to be able to distinguish between five models, which were chosen to lie along the $\sigma _8$ - $\Omega _m$ degeneracy, and have nearly the same two-point statistics.
We design and implement a Deep Convolutional Neural Network (DCNN) which learns the relation between five cosmological models and the mass maps they generate.
We develop a new training strategy which ensures the good performance of the network for high levels of noise.
We compare the performance of this approach to commonly used non-Gaussian statistics, namely the skewness and kurtosis of the convergence maps.
We find that our implementation of DCNN outperforms the skewness and kurtosis statistics, especially for high noise levels.
The network maintains the mean discrimination efficiency greater than $85\%$ even for noise levels corresponding to ground based lensing observations, while the other statistics perform worse in this setting, achieving efficiency less than $70\%$.
This demonstrates the ability of CNN-based methods to efficiently break the $\sigma _8$ - $\Omega _m$ degeneracy with weak lensing mass maps alone.
We discuss the potential of this method to be applied to the analysis of real weak lensing data and other datasets.
}

\begin{document}
\maketitle
\flushbottom

\newpage


\section{Introduction}
\label{sec:intro}

Recent observations have indicate that the density of matter in the Universe evolves with cosmic time \citep[see, for example, ][]{springel2005simulations, navarro1995structure, bond1996filaments}. Starting from initial Gaussian fluctuations, the matter density field becomes increasingly non-linear, and gives rise to non-Gaussian features, such as halos, filaments and sheets \citep{hahn2006properties}. Statistics of this large scale distribution of matter in the Universe are widely used to constrain the parameters of cosmological models. Weak gravitational lensing measurements provide a powerful dataset that can be used for measuring these statistics \citep{bartelmann2001weak,kilbinger2014cosmology,hildebrandt2017kids,des2015cosmology}.
Other probes are also used for this purpose, for example galaxy clustering \citep{hutsi2006power,beutler2014clustering} and clusters \citep{rozo2009cosmological,vikhlinin2008chandra}, among others.

\begin{figure}\centering
\includegraphics[width=0.5\textwidth]{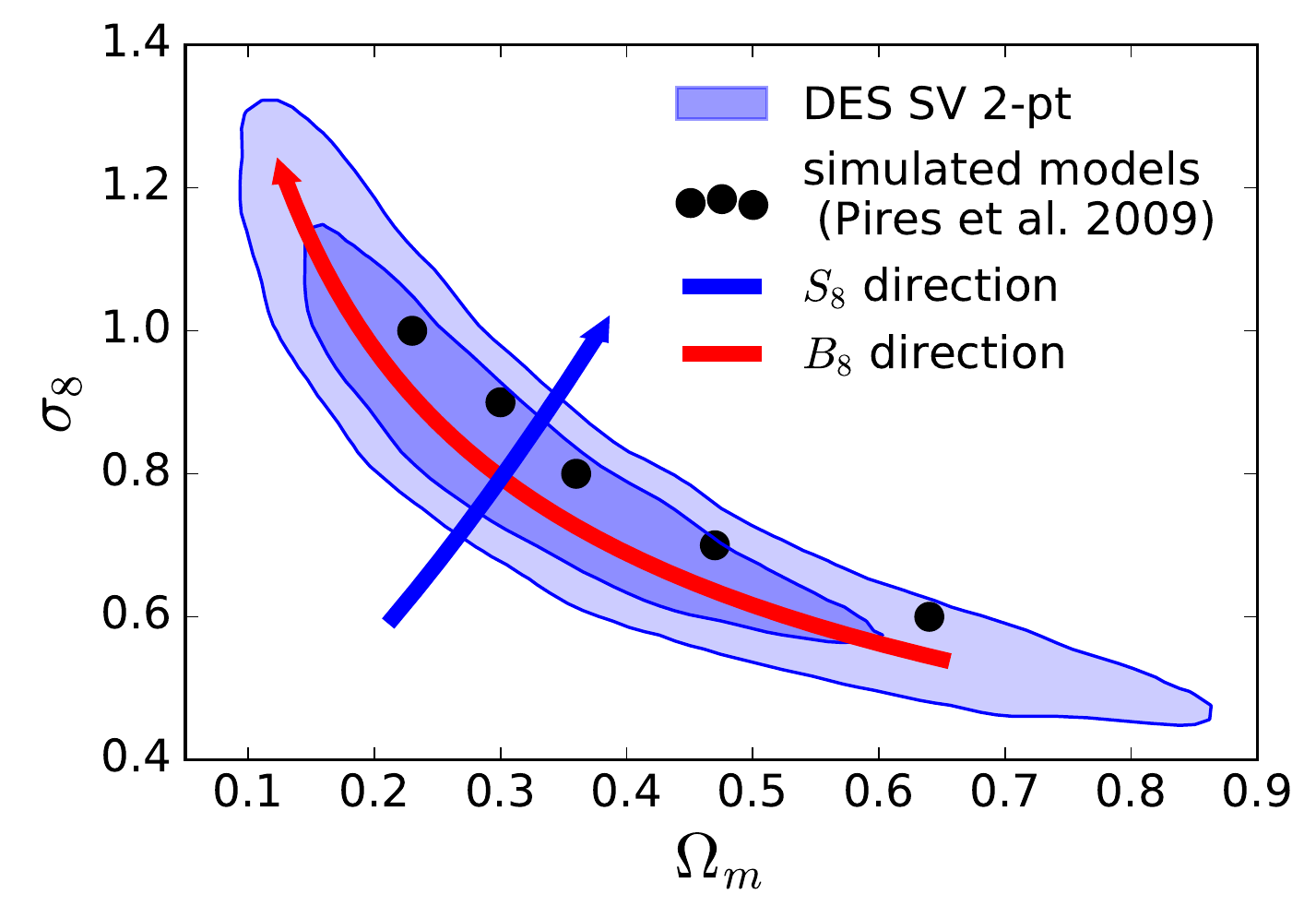}
\caption{Constraints on $\Omega_m$ and $\sigma_8$ from weak lensing analysis of \citep{des2015cosmology}. The blue line shows the direction across the degeneracy, defined as $S_8=\sigma_8 (\Omega_m/0.3)^{0.6}$. This direction can be efficiently constrained by 2-pt statistics. The direction along the degeneracy, $B_8$, is marked with a red line. Black points show the five models from \citep{pires2009model}, which were also used in this work. The values of parameters for these models are shown in Table~\ref{tab:5cosmparams}.  }
\label{fig:banana}
\end{figure}

The methods used most commonly for these studies are two-point statistics, namely the power spectrum, or its real space equivalent, the correlation function \cite{peacock1991power,schneider2002analysis,kilbinger2014cosmology}. For Gaussian random fields, these statistics capture all the available information. However, as discussed above, the fields of interest in cosmology can be highly non-Gaussian and thus contain information beyond the 2-pt function. Various cosmological models give rise to density fields that differ in structure, and yet have similar two-points statistics \citep{hildebrandt2017kids,des2015cosmology}.  This leads, in particular, to a degeneracy in the weak lensing constraint on the matter density $\Omega_m$ and power spectrum normalisation $\sigma_8$. Recent constraints on $\Omega_m$ and $\sigma_8$ from the DES collaboration\footnote{\url{www.darkenergysurvey.org}} \citep{des2015cosmology} are shown in Figure~\ref{fig:banana}. The direction across the degeneracy, marked with a blue arrow and often referred to as $S_8$, is well constrained by the cosmic shear power spectrum. The power spectrum, however, cannot strongly constrain the parameters along the degeneracy direction, marked with a red arrow, which we will refer as $B_8= ((1-(\Omega_m))/0.3)^{0.6}/(3-\sigma_8)$ (see Section \ref{sec:simulations}). Other recent weak lensing analyses show similar degeneracy \citep{hildebrandt2017kids,vanUitert2017kidsgama}.
A range of methods have thus been proposed with the aim of breaking that degeneracy using weak lensing alone. In order to do so, a method must be able to efficiently extract information beyond Gaussian.

Bi-spectrum or three-point correlation function, is commonly used to capture high-order moments of the matter density distribution \citep{sefusatti2006cosmology,dodelson2005weak,fu2014cfhtlens}. Other statistics, such as skewness and kurtosis \citep{takada2002kurtosis,vafaei2009breaking} have been studied. Peak statistics \citep{kacprzak2016cosmology,liu2014cosmology,dietrich2009cosmology}, which simply count peaks in the reconstructed density maps, are sensitive to high density regions of the distribution. Minkowski functionals \citep{pratten2012nongaussianity,kratochvil2011probing,matilla2017geometry} have been also proposed to capture the non-Gaussian components in density maps.  A study was conducted in \citep{pires2009model} in order to assess the ability of a variety of high-order statistics to discriminate between cosmological models with very similar power spectrum. The performance of the different methods considered varied, with some reaching good discrimination precision in the degeneracy direction.

In order to make a prediction from a summary statistic for a range of cosmological parameters, it is common to calculate statistics directly from simulations, which forward-model the matter density evolution. Therefore, a method to discriminate between cosmological models can be freely designed, as long as it is able to treat the simulated and observed data consistently. In this work, we explore the possibilities of using Deep Convolutional Neural Networks (DCNN) \citep{George2016a,Dieleman2015a} as a method for discriminating between cosmological models. This type of neural network has been shown to achieve high discrimination performance for natural images \citep{Krizhevsky2012a}. Given enough training data, DCNNs can learn  a complex set of filter functions that, when applied to images in a hierarchical fashion, maximise the discrimination performance. This alleviates the need of having to manually define image filters. Deep networks have also recently gained more attention for astronomy applications including for the prediction of galaxy morphology~\cite{Dieleman2015a} or generative models for images of galaxies~\cite{ravanbakhsh2016enabling}.

We design a DCNN based classification system to discriminate between cosmological models using the projected matter density distribution. We follow the methodology in \citep{pires2009model}, where the model discrimination problem was simplified: the methods aim to distinguish between several, discretely sampled cosmological models. We compare the performance of the DCNN approach to two commonly used high-order statistics, namely the skewness and kurtosis. We consider practical cases where noise degrades the mass maps and perform comparisons for varying levels of noise.

The paper is organised as follows. We start by describing the process of simulation data used in this work in Section~\ref{sec:simulations}, and explain the process used to create weak lensing convergence mass maps in Appendix~\ref{sec:convergence_map_calculation}. We then present a short overview of related work and alternative statistical methods for discriminating different convergence maps in Section~\ref{sec:methods}. An introduction to Convolutional Neural Networks is given in Section~\ref{sec:deep_network}, as well as the details of the architecture and training procedure used for our network. We present our results in Section~\ref{sec:results}. An analysis of the training process and learned image filters is performed in Section~\ref{sec:network_analysis}. We conclude and give an outlook to possible extensions of this work in Section~\ref{sec:summary}.


\section{Simulations}
\label{sec:simulations}

\begin{table}
\centering
	\begin{tabular}{lccccccc}
	\hline
	Model	&$L$		&$\Omega _m$	&$\Omega_{\Lambda}$	&$h$	&$\sigma _8$	&$\Omega _B$	&$n_s$ \\ \hline
	model 1	&161.17		&0.23			&0.77			    &0.594	&1.0			&0.04			&0.958	\\
	model 2	&154.18		&0.30			&0.70			    &0.700	&0.9			&0.04			&0.958	\\
	model 3	&149.05		&0.36			&0.64			    &0.798	&0.8			&0.04			&0.958	\\
	model 4	&141.12		&0.47			&0.53			    &0.894	&0.7			&0.04			&0.958	\\
	model 5	&131.48		&0.64			&0.36			    &0.982	&0.6			&0.04			&0.958	\\
	\hline
	\end{tabular}

	\caption{Cosmological parameters of the five models along the $\Omega_m - \sigma_8$ degeneracy. $L$ is the box length in $\mathrm{Mpc} h ^{-1}$. For more details of the simulations, see Appendix~\ref{sec:convergence_map_calculation}.}
	\label{tab:5cosmparams}
\end{table}

In the early universe, the matter density distribution can be well described by a Gaussian Random field, but its evolution throughout cosmic time caused it to become increasingly non-Gaussian, displaying non-linear features such as filaments, halos and sheets. This distribution is commonly called the ``Cosmic Web''. Simulations of the evolution of matter density are commonly conducted using a N-Body techniques \citep[see, for example,][and references therein]{cautun2014evolution,springel2005simulations,springel2003gadget,potter2016pkdgrav3}. The three-dimensional distribution of matter particles can then be projected onto a two-dimensional, weak lensing convergence plane.
In this section, we describe the simulation process used to generate two-dimensional patches of matter distribution for various configurations of cosmological parameters. These simulations are then used to train our DCNN approach and evaluate its ability, as well as that of other statistical methods, to discriminate between cosmological models.

In order to simulate the evolution of matter density, we use the $\Lambda$CDM cosmological model, which depends on the following parameters:
(i) $\Omega_m $ the total matter density,
(ii) $\sigma _8$, the mass fluctuation amplitude on $8 \ \mathrm{Mpc} h^{-1}$ scales.
Variations of these parameters will manifest in a different composition and local densities of the universe and consecutively in different statistics retrieved from the mass maps of these densities.
All other parameters in the  $\Lambda$CDM model have been fixed, such as, for example: the Hubble constant $H_0$, representing the expansion rate of the universe,
$\Omega_b$, the density of the baryonic matter, and $n_s$, the scalar spectral index, which describes the variation of density fluctuations with scale.
We assume no spatial curvature and set the dark energy density $~\Omega_{\Lambda} = 1-\Omega_m$.

\begin{figure}
\includegraphics[width=1.0\textwidth]{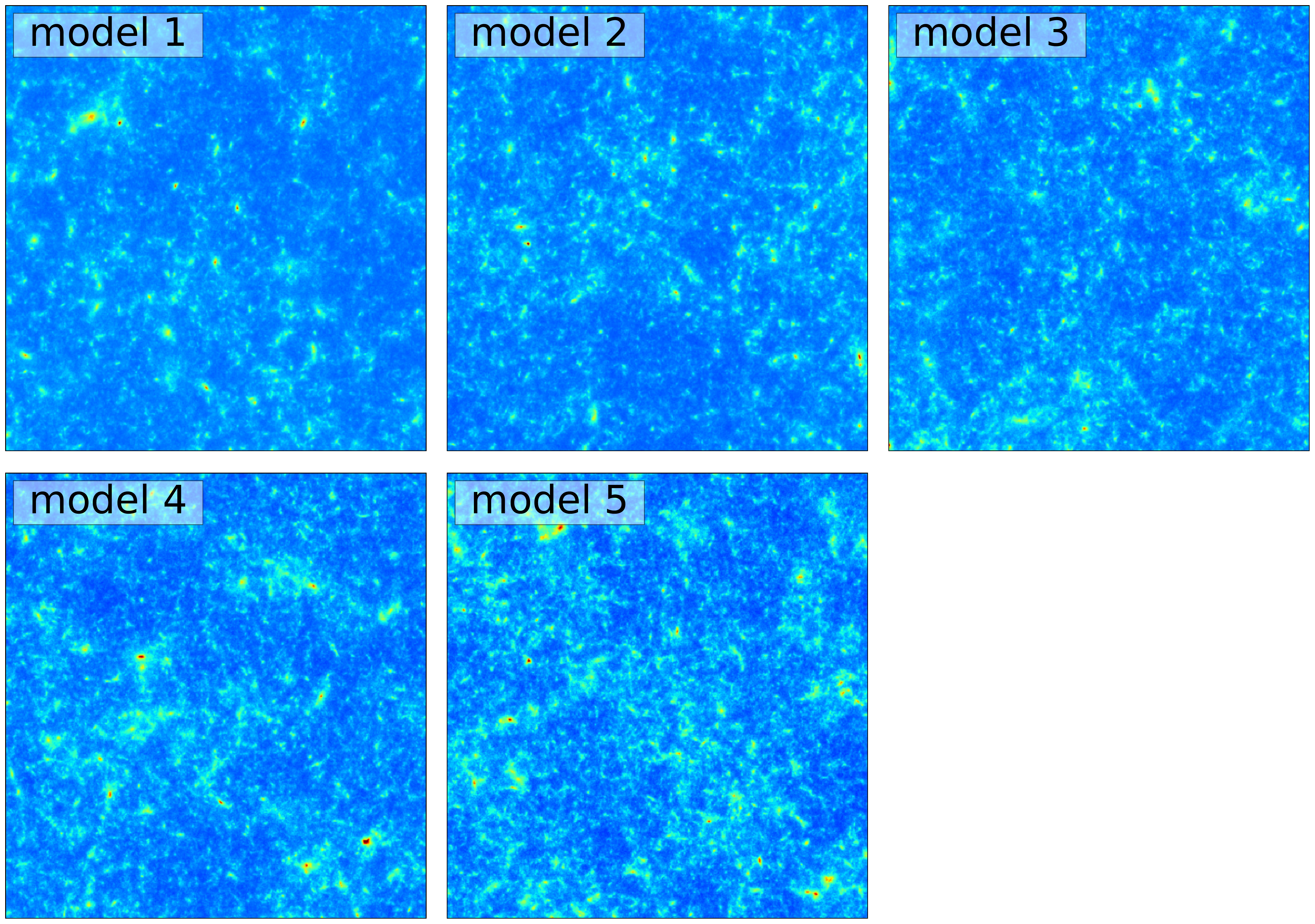}

\caption{Examples of noiseless convergence maps for the five different cosmologies listed in Table~\ref{tab:5cosmparams}. The field size is downsampled to $512 \times 512$ pixels, which corresponds to $ 1.9^{\circ} \times 1.9^{\circ} $ in our case and is one fourth of the original map size in terms of number of pixels. \\}
\label{fig:example_maps}
\includegraphics[width=1.0\textwidth]{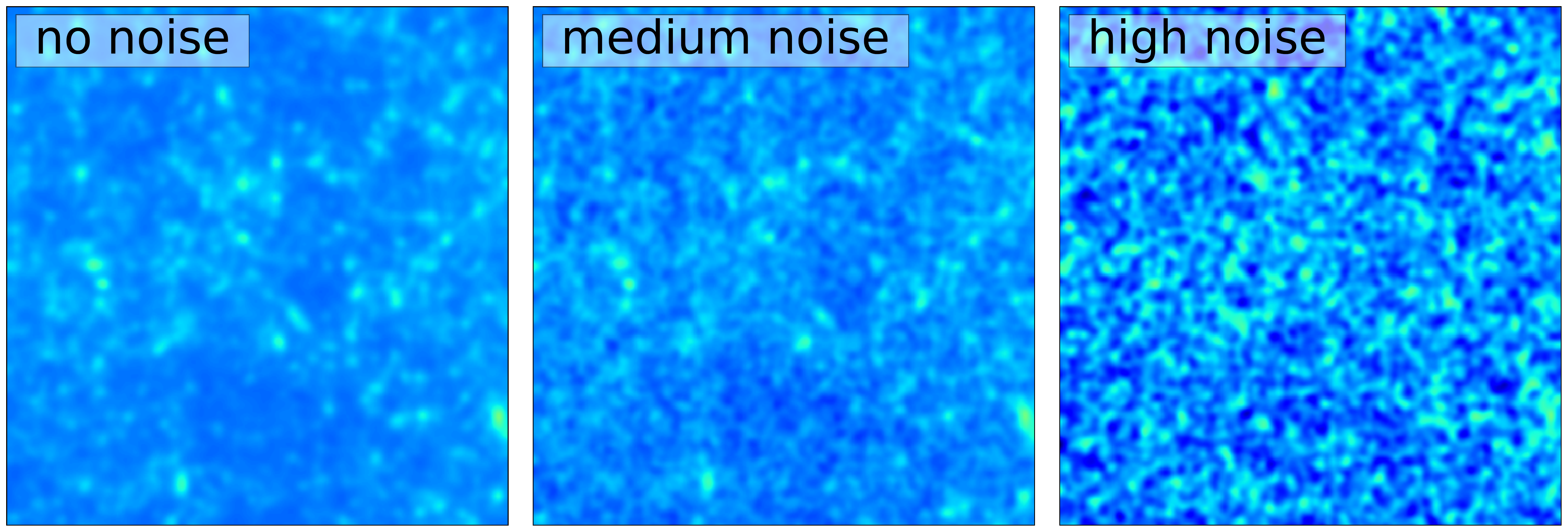}
\caption{Example maps from Model 2 with varying level of noise added: noiseless (left), moderate noise corresponding to space based observations with number of galaxies $n_{g}=100 / \rm{arcmin}^2$ (middle) and high noise corresponding to ground based observations with number of galaxies $n_{g}=10 / \rm{arcmin}^2$ (right). The maps were smoothed with a kernel of size $\sigma_{\rm{smoothing}}=0.9 \ \rm{arcmin}$ (see Section~\ref{sec:methods}).}
\label{fig:example_maps_noisy}
\end{figure}

We follow the procedure described in \citep{pires2009model} to generate the data simulation used for both training and evaluation. In a first step, we produce mock maps for five different cosmological models along the $\sigma _8 - \Omega _m$ degeneracy, as done in \citep{pires2009model}. The respective cosmological parameters of the different realizations of the $\Lambda$CDM cosmologies are given in Table \ref{tab:5cosmparams}.
To simulate cosmic structures, we used a fast simulation code, \textsc{L-PICOLA} \citep{Howlett2015a}, that evolves a 3D mass distribution in cosmic time. We then project the 3D distribution onto a 2D sky plane, following \citep{Vale2003a,pires2009model}.
For each cosmological model, we create 2500 independent convergence maps of size $ 1.9^{\circ} \times 1.9^{\circ} $. The details of this procedure is described in Appendix~\ref{sec:convergence_map_calculation}.

The levels of noise added to these maps correspond to the levels expected from a space-survey with $n_g^{\rm{space}}=100$ galaxies/arcmin$^2$, and ground-based survey with $n_g^{\rm{ground}}=10$ galaxies/arcmin$^2$. More details on the noise simulation can be found in Appendix~\ref{sec:convergence_map_calculation}.

Examples of noiseless maps created from five cosmological models are presented in Figure~\ref{fig:example_maps}.
Examples of convergence maps with added noise and smoothing (see Section~\ref{sec:methods}) are shown in Figure~\ref{fig:example_maps_noisy}.


\section{Discrimination method}
\label{sec:methods}
\label{sec:classifier_for_skew_kurtosis}

Following \citep{pires2009model}, we compare the performance of the DCNN to two methods using the skewness \sskew \ and kurtosis \kkurt \ statistics computed on the convergence maps. Each of these methods has an ability to classify a new mass map between any pair of models $i$ and $j$.
We perform a discrimination of all maps created with true model $i$ against all models $j \neq i$ (the discrimination of a model against itself is not meaningful and is thus not reported). Given our five cosmological models, each method can thus perform $5 \cdot 4$ classifications. The discrimination efficiency is summarised in a \emph{Confusion Matrix} (CM). An entry $\mathrm{CM}_{ij}$ is defined as a percentage of maps with true model $i$ classified correctly against a model $j$.

\begin{figure}
\centering
\includegraphics[width=0.4\textwidth]{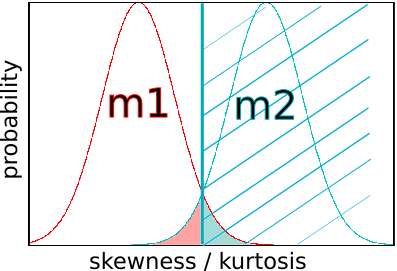}
\caption{Demonstration of the construction of the decision boundary between two classes.
By solving the equation $Q_{1-x} \kll m_1 \klr = Q_{x} \kll m_2 \klr$, we find the decision boundary which maximizes classification accuracy for skewness- and kurtosis-based methods.
The red and cyan lines show the distributions of a statistic computed from multiple instances of mass maps, for models m1 and m2, correspondingly.
The vertical thick cyan line shows the calculated boundary.
All maps, for which the values of a statistic is greater than the boundary, will be classified as model m2 (cyan hashed region).
}
\label{fig:new_discrimination}
\end{figure}

For the skewness and kurtosis methods, we construct the following classifier. For each cosmological model and noise level, we aim to create a probability distributions $p(\sskew)$ and $p(\kkurt)$ calculated from single $1.9^{\circ} \times 1.9^{\circ}$ mass maps. We create this distribution using histograms of \sskew \ and \kkurt \ calculated from the full set of mass maps.
These histograms are then used to create a set of decision boundaries between each pair of cosmological models, as illustrated in Figure~\ref{fig:new_discrimination}.
We find the separating boundary between the two class distributions by equating the number of false positives of the two distributions under the assumption that the distributions contain the same number of samples. Formally, let $Q_x(D)$ be the $x \times 100 \%$-Quantile of distribution $D$, we then need to solve the equation $ Q_{1-x} \kll m_1 \klr = Q_{x} \kll m_2 \klr $ for $x$, which is the classification boundary.

When using noisy convergence maps, we additionally smooth them before calculating the skewness and kurtosis, which helps to reduce the impact of the noise. We used a Gaussian as a smoothing kernel whose width was optimised to maximise the discriminating power of the method. We found that a standard deviation of $\sigma_{\rm{smooth}}=0.9 \ \rm{arcmin}$ reported in \citep{pires2009model} performed the best. We also found that there was no need to vary $\sigma_{\rm{smooth}}$ according to the noise level and thus kept this value fixed for all cases.
The next section will describe how the neural network discriminates between a pair of models.

\section{Discrimination using Deep Learning}
\label{sec:deep_network}

One of the advantages of Deep Neural Networks is their ability to automatically learn the right data representation without the need to design a set of statistics \emph{a priori}. We here briefly review how they operate and direct the reader to \citep{Goodfellow2016} for further details.

\subsection{Nomenclature and functionality of Neural Networks}
\label{sec:deep_network_nomenclature}

\paragraph{Neural network model.} A neural network consists of a complex combination of neurons organized in nested layers. Each neuron implements a function that is parametrized by a set of weights $\w \in \mathbb{R}^d$. Every layer of a neural network thus transforms one input tensor to another through a differentiable function.
Formally, given a neuron $n$ receiving an input vector ${\bf x_n} = (x_n^{1}, \dots, x_n^d) \in \mathbb{R}^d$, and the choice of an activation function $f_n$~\footnote{The typical role of an activation function is to make neural networks non-linear.}, the output of the neuron is computed as
\begin{align}
	o_n = f_n \kll \sum_{i=1}^d w_n^i \cdot x_n^i \klr
\end{align}
where ${\bf w_n} = (w_n^1, \dots, w_n^d)$ are the parameters of the neuron $n$. Common choices for the activation function $f_n$ include the sigmoid, $\tanh$ functions or rectified linear units (ReLUs) define as $f_n(\cdot) = \max(0, \cdot)$.

The combination of neurons organized in layers create a model or function $f: x \to y$ from an input image $\x$ to an output label $y \in \{1, \dots 6 \}$. This function is parametrized by the weights encoded in each layer. We will here denote by $\mathbf W^{l}$ the weights contained in the $l$-th layer, i.e. for a layer $l$ containing $m$ neurons, $\mathbf W^{l} = (\w_1, \dots, \w_m)^\top$. The entire model represented by the neural network $f$ is thus parametrized by $\W = (\mathbf W^1, \dots, \mathbf W^{L})$.\\

In this work we use a special type of neural network known as \emph{Convolutional Neural Networks} (CNNs), which have been empirically shown to perform well for various image tasks. One task where they have particularly excelled is image classification \citep{Krizhevsky2012a} where they achieved results near human accuracies \citep{simonyan2014very}. Three main types of layers are used to build a CNN architecture: Convolutional Layer, Pooling Layer, and Fully-Connected Layer~\footnote{These layers are sometimes called hidden layers.}. Note that some layers contain parameters and other do not. The role of these layers will be explained in more details in Section~\ref{sec:details_CNN}.

\paragraph{Training a CNN.} The weights of a CNN are learned from data samples in a supervised manner. Provided a training data set $\mathcal{X} = \{ (\x_t, y_t): t=1, \dots, T \}$ consisting of input samples $x_t$ and the corresponding \emph{ground truth} $y_t$ (i.e. the desired output of the network), we first define a loss function $\loss(y, \ybar)$ between the ground truth $y$ and the output of the CNN $\ybar$. A common choice for $\loss$ is the Cross-entropy loss, which in the case of binary random variables $y \in \{0, 1\},\ybar \in ]0, 1[$ is defined as
\begin{align}
\loss(\ybar; y) = - y \log \ybar - (1- y) \log (1-\ybar).
\end{align}
The best set of weights $\W^*$ are then obtained by minimizing an empirical risk function,
\begin{align}
\W^* = \arg \min_{\W} \left[ \Loss(\W; \mathcal X) = \frac 1 T \sum_{t=1}^T \loss(y_t; \ybar), \right]
\quad \W = (\mathbf W^1, \dots, \mathbf W^{L})
\label{eq:emp_loss}
\end{align}
where $\ybar=y(\x_t; \W)$ is the output of the neural network for point $\x_t$ and weights $\W$. Minimizing Eq.~\ref{eq:emp_loss} is typically done using stochastic gradient descent (SGD) and thus requires the computation of the gradient through the layers of the neural network, which is made possible by using the chain rule to iteratively compute gradients for each layer. This is implemented using the back-propagation algorithm.
The performance of the trained network is assessed by evaluating its accuracy, which measures the number of samples that are correctly classified. While the accuracy computed on the training data is used to evaluate the choice of parameters, we also compute the accuracy on a separate test set.

Note that $L_2$ regularization or weight decay is commonly used to avoid \emph{overfitting} to the training set. A modern variant is known as dropout \citep{Srivastava2014} which proceeds as follows. At each step of the training procedure a given neuron is selected with a predefined probability $1-p_{keep}$. The selected neurons act as if they were not part of the network and the corresponding weights are thus not updated at this iteration. This prevents the network from overtraining on specific examples and results in better generalization properties. At test time we use all weights and set to $p_{keep} =1$.

\subsection{Basic elements of a CNN}
\label{sec:details_CNN}

As illustrated in Figure~\ref{fig:architecture}, a CNN processes an input image by converting the original pixel values to the final class scores. This process is performed by three types of layers that we will now briefly describe.

\paragraph{Convolutional layers. } These layers take advantage of the strong spatial correlations in images by swiping and convolving different learned \emph{filters} $\rm{F}$ over the image (input $I$). Each filter detects a specific pattern in the image and produces a feature map $FM$ using a convolution, i.e.
\begin{align}
\rm{FM}\kll i,j\klr = \kll I\star F\klr \kll i,j \klr := \sum _k \sum _l I \kll i+k ,j+l \klr F\kll k,l\klr.
\end{align}
The resulting feature maps are typically processed by an activation function (e.g. a ReLU) before they eventually undergo further convolutions or other operations depending on the network architecture.

\paragraph{Pooling layers. } They reduce the size of their input, typically by performing a down-sampling operation. They can also be performed using a convolution with a \emph{stride} of 2 (a stride is the distance between two consecutive positions of the filter).

\paragraph{Fully connected layers. } They perform the high-level reasoning by connecting all neurons in the previous layer to every single neuron it
has. This operation can typically be computed as a matrix multiplication. If used at the last layer in the network, it outputs a set of scores, one for each class (five in Figure~\ref{fig:architecture}).

\subsection{Architecture}
\label{sec:architecture}

\begin{figure}
\centering
\includegraphics[width=0.9\textwidth]{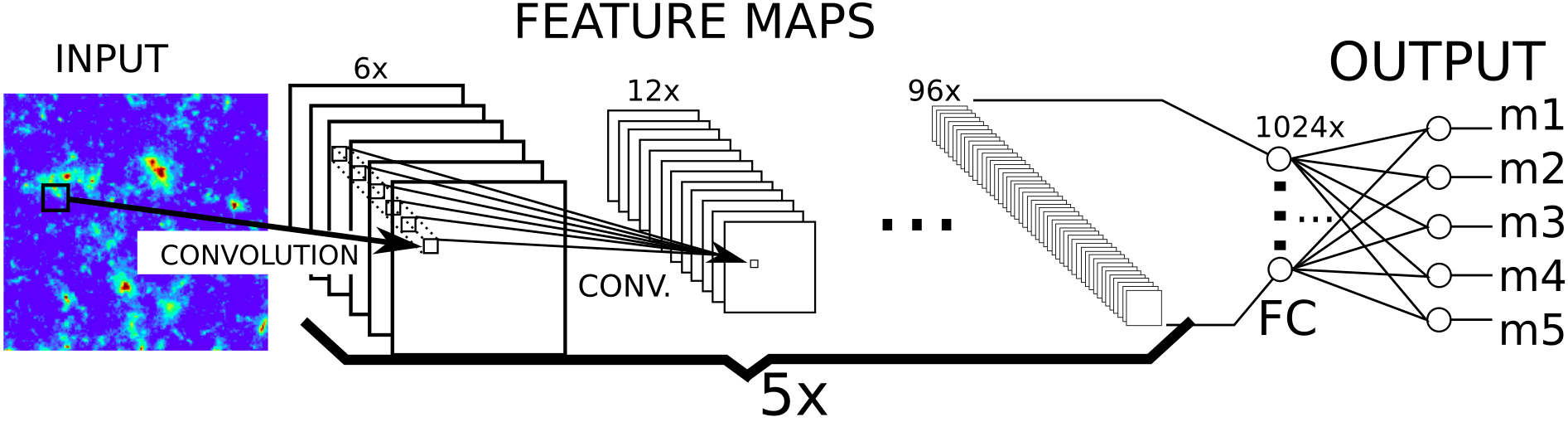}
\caption{Sketch of the DCNN architecture used for discriminating between convergence maps. A sequence of convolutional layers with an increasing number of filters as explained in Section~\ref{sec:architecture} is followed by a fully connected layer (FC) containing 1024 neurons. The last element of the network is a fully connected output layer.}
\label{fig:architecture}
\end{figure}

The choice of an architecture can have an important influence on the performance of the network. Several design decisions have to made concerning the number and the type of layers, as well as the number and the size of the filters used in each layer. The best way to make these choices is typically through experimentation although some guidelines can be found in the literature \citep{Goodfellow2016}. This includes the size of the network, which depends on the number of training examples as networks with a large number of parameters are likely to overfit if not enough training examples are available. For our application we found that common deep architectures used in image recognition, e.g. VGG-16 \citep{simonyan2014very} were not well-adapted due to the limited amount of training data. Note that more data could be produced from simulations but we instead opted for using a slightly less complex network architecture, named DCNN, which is presented in Figure~\ref{fig:architecture}. Following the input layer, the network contains 6 hidden layers including a fully connected output layer. The first five hidden layers perform convolutions while the fully connected layer contains 1024 ReLU neurons with dropout regularization. The dropout probability is $p_{\rm{keep}} = 0.5$ during training (and $p_{\rm{keep}} = 1$ at test time).

In the first convolutional layer, we convolve our input with 6 different filters, each of size $5 \times 5$ pixels and using a stride of 2. The output of this convolutional layer consists of 6 feature maps. In each following convolutional layer, the number of filters of the preceding layer is multiplied by 2, while the filter size and stride are kept constant. This amounts to a total of around 106.1 million trainable parameters in our network compared to 138 million trainable parameters for the VGG-16 network.

\subsection{Training strategy}
\label{sec:strategy}

The objective function of neural networks is known to have many saddle points and local minima, which makes the optimization process very difficult \citep{Swirszcz2016a, Kawaguchi2016a}. In real-world scenarios, high levels of noise degrade the training data and typically results in optimization landscapes with more local minima and therefore increases the difficulty in training the neural network. It can thus be desirable to start optimizing the neural network using noise-free data which typically yield smoother landscapes. This type of approach is commonly known as continuation methods or curriculum learning~\cite{bengio2009curriculum}.  The basic idea is to define a sequence of objectives $f(\sigma_t)$ over which we optimize such that the sequence approaches some desired final target function $f(\sigma_\text{target})$. Energy based continuation methods, such as, for example, simulated annealing, have also been commonly used in physics, computer vision \citep{blake1987visual} and astrophysics\citep{George2016a}.

Based on this observation, we thus defined a new training strategy designed to help the network to adapt to increasing levels of noise by using the network parameters obtained from less noisy data. Three variations of this strategy were explored (called \emph{DCNN}, \emph{DCNNf} and \emph{DCNNr}) depending on the increase used to level up the noise level during training. Details of these variations will follow after a broad description of the algorithm.

The noise-adaptive training procedure is summarized in Algorithm~\ref{alg:training_strategy} and is somewhat similar to an online data augmentation approach. At each step of training, noise is added to the original training data according to a current noise level $\sigma_{\kappa}$. This is performed by creating new images that are noisy versions of the original training images, effectively increasing the size of the training set. This noise level is increased when the training accuracy goes above a fixed threshold $T_{a}$ whose value depends on the choice of strategy.  A new random seed is fixed for each specific noise level except for high levels of noise (greater than $\sigma_{\kappa}=0.2+/-0.05$) for which we have observed that the network had difficulties coping with the high variance of the augmented noisy data. We thus did not re-initialize the seed passed the threshold  $\sigma_{\kappa}=0.2+/-0.05$.

\begin{table}[h]
\begin{tabular}{l}
\hline
\begin{algorithm}[H]
 \KwData{Noiseless weak lensing convergence mass maps}
 \KwResult{DCNN capable of discriminating between given cosmological models in noisy weak lensing mass maps with noise standard deviation $\sigma _{\kappa}^{\rm{ground}}$}
 $\sigma_{\kappa} = 0$\;
 $\Delta\sigma_{\kappa} = 0.003$\;
 $\sigma_{\kappa}^{\rm{ground}} = 0.42$\;
 \While{$\sigma_{\kappa} < \sigma_{\kappa}^{\rm{ground}}$}{
  $\mathcal B \leftarrow $ Sample mini-batch of data \\
  DCNN training step $\kll \mathcal B, \sigma_{\kappa}\klr$ \\
  $\alpha \leftarrow$ Compute training accuracy over $\mathcal B$ \\
  \eIf{$\alpha > T_{a}$ \% for three consecutive steps}{
   $\sigma_{\kappa} += \Delta\sigma_{\kappa}$ \\
   reset \textsc{Adam} optimizer
   }{
   continue
  }
 }
 \caption{Noise adaptive strategy used to train the DCNNs. Once the network has been pre-trained on noiseless data, noise levels were gradually increased by $\Delta\sigma_{\kappa} = 0.003$ every time the network accuracy - over three consecutive training steps - went above a pre-defined threshold $T_{a}$.}
 \label{alg:training_strategy}
\end{algorithm}\\
\hline
\end{tabular}
\end{table}

As mentioned previously, we implemented three different strategies to increase the noise level, namely
\begin{enumerate}
\item [\emph{DCNN}] (our main result): Algorithm~\ref{alg:training_strategy} with $T_{a} = 90 \%$ after pre-training on noiseless maps and did not change it any more.
\item [\emph{DCNNf}] (fast) the noise is levelled up with $\Delta \sigma _{\kappa} = 0.01$ at each iteration, i.e., unlike the \emph{DCNN} strategy, the level of noise is increased irrespective of the accuracy level. Once the noise level $\sigma _{\kappa} =\sigma _{\kappa}^{space}= 0.13$ is reached, then we switch to the \emph{DCNN} strategy.
\item [\emph{DCNNr}] (faster): Algorithm~\ref{alg:training_strategy} with a threshold $T_{a} = 80 \%$ after pre-training on noiseless maps.
\end{enumerate}

We optimize each network using a variant of SGD known as \textsc{Adam} \citep{Kingma2014a} with a learning rate of $10^{-4}$ and hyper-parameters $\beta_1=0.9$ and $\beta_2=0.999$. The total time to train the neural network was around 500 hours on a machine equipped with an Nvidia Tesla P100 GPU with double-precision performance of around 5 teraflops.

\section{Results}
\label{sec:results}

In order to evaluate the performance of the DCNN approach and compare it to skewness and kurtosis statistics, we compute the discrimination efficiency between each possible pair of models for all approaches and for varying noise levels. We summarize the results of all pairwise comparisons in a single CM. To summarise the results of all pairwise comparisons as one number, we compute the mean discrimination efficiency $\rm{MDE}$, defined as the mean of the entries of the confusion matrix, neglecting the diagonal, i.e.
\begin{equation}
    \mathrm{MDE}\kll \mathrm{CM} \klr = \frac{1}{N^2-N}\sum _{i,j} ^N \kll 1-\delta _{ij}\klr \mathrm{CM}_{ij},
\end{equation}
where $N = 5$ is the number of cosmological models.
Note that the $\rm{MDE}$ score scales between $[0, 100]$ since each entry in a given confusion matrix $\mathrm{CM}_{ij}$ is assumed to be a percentage. A result equal to $ \leq 50 \%$ corresponds to random guessing and we therefore do not expect a score to be lower than $50 \%$.\\

\begin{figure}
\centering
\includegraphics[width=0.9\textwidth]{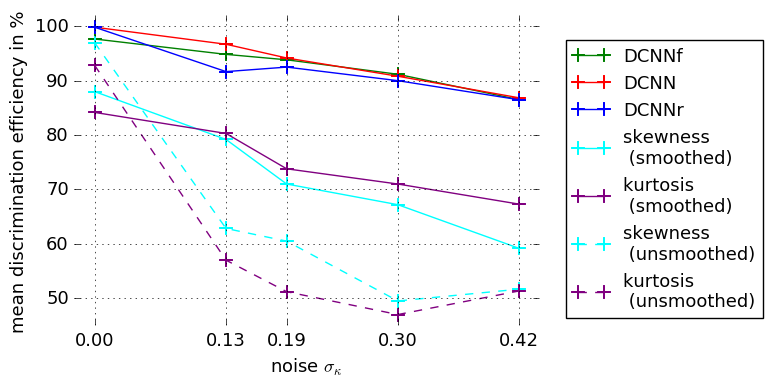}
\caption{Mean discrimination accuracy of DCNN compared to the skewness and kurtosis statistics, as a function of the noise level added to the convergence maps. Before calculation of the skewness and kurtosis statistics, the maps were smoothed with a filter of size $\sigma_{\rm{smoothing}}=0.9 \ \rm{arcmin}$. Results from the maps without smoothing are also shown.}
\label{fig:mean_discrimination_efficiency}
\end{figure}

\begin{table}[h]
\scriptsize
\begin{centering}
\begin{tabular}{c@{\hskip 0.12in} c@{\hskip 0.12in} c@{\hskip 0.12in}}
\ & {\bf no noise}  & \ \\
\hline
skewness & kurtosis  & DCNN \\
\hline
\\
  \begin{tabular}{|c@{\hskip 0.12in} c@{\hskip 0.12in} c@{\hskip 0.12in} c@{\hskip 0.12in} c@{\hskip 0.12in}|}
    \	& 90.0	 & 96.7	 & 100	 & 100	 \\
    90.0	&  \ & 70.0  & 90.0	 & 100	 \\
    90.0 & 66.7	 & \	 & 66.7  & 100	 \\
    93.3 & 90.0	 & 86.7  & \	 & 70.0 \\
    96.7 & 93.3  & 90.0	 & 80.0  & \  \\
\end{tabular}
&
  \begin{tabular}{|c@{\hskip 0.12in} c@{\hskip 0.12in} c@{\hskip 0.12in} c@{\hskip 0.12in} c@{\hskip 0.12in}|}
    \   & 86.7   & 93.3  & 100   & 100   \\
    80.0    &  \     & 63.3 & 80.0   & 96.7  \\
    86.7 & 56.7  & \     & 76.7 & 86.7   \\
    93.3 & 90.0      & 76.7 & \  & 63.3 \\
    93.3 & 90.0      & 90.0  & 80.0 & \\
\end{tabular}
&
\begin{tabular}{|c@{\hskip 0.12in} c@{\hskip 0.12in} c@{\hskip 0.12in} c@{\hskip 0.12in} c@{\hskip 0.12in}|}
    \   & 100    & 100   & 100   & 100   \\
    100 & \      & 100   & 100   & 100   \\
    100 & 96.7   & \     & 100   & 100   \\
    100 & 100    & 100   & \     & 100 \\
    100 & 100    & 100   & 100   & \\
\end{tabular}
\\
\vspace{5 mm}
\\ & {\bf moderate noise}  & \ \\
\hline
skewness & kurtosis  & DCNN \\
\hline
\\
  \begin{tabular}{|c@{\hskip 0.12in} c@{\hskip 0.12in} c@{\hskip 0.12in} c@{\hskip 0.12in} c@{\hskip 0.12in}|}
    \   & 76.7  & 83.3     & 90.0     & 96.7     \\
    76.7   & \      & 60.0 & 76.7     & 93.3     \\
    80.0 & 60.0   & \     & 60.0 & 80.0    \\
    93.3 & 86.7      & 80.0 & \    & 60.0 \\
    90.0 & 86.7    & 83.3    & 70.0 & \\
\end{tabular}
&
  \begin{tabular}{|c@{\hskip 0.12in} c@{\hskip 0.12in} c@{\hskip 0.12in} c@{\hskip 0.12in} c@{\hskip 0.12in}|}
    \   & 83.3  & 86.7     & 90.0     & 100     \\
    76.7   & \      & 60.0 & 80.0     & 83.3     \\
    80.0 & 53.3   & \     & 70.0 & 86.7    \\
    93.3 & 90.0      & 73.3 & \    & 60.0 \\
    90.0 & 90.0    & 83.3    & 76.7 & \\
\end{tabular}
&
\begin{tabular}{|c@{\hskip 0.12in} c@{\hskip 0.12in} c@{\hskip 0.12in} c@{\hskip 0.12in} c@{\hskip 0.12in}|}
    \   & 100    & 100   & 100   & 100   \\
    100 & \      & 63.3   & 100      & 100   \\
    100 & 73.3   & \     & 96.7  & 100   \\
    100 & 100    & 100   & \     & 100 \\
    100 & 100    & 100   & 100   & \\
\end{tabular}
\\
\vspace{5 mm}
\\ & {\bf high noise}   & \ \\
\hline
skewness & kurtosis  & DCNN \\
\hline
\\
  \begin{tabular}{|c@{\hskip 0.12in} c@{\hskip 0.12in} c@{\hskip 0.12in} c@{\hskip 0.12in} c@{\hskip 0.12in}|}
    {\   } & {56.7}& {73.3} & {73.3} & {30.0} \\
    {80.0} & {\   }& {63.3} & {63.3} & {56.7} \\
    {70.0} & {66.7}& {\   } & {53.3} & {60.0} \\
    {76.7} & {66.7}& {36.7} & {\   } & {66.7} \\
    {26.7} & {43.3}& {63.3} & {56.7} & {\   } \\
\end{tabular}
&
  \begin{tabular}{|c@{\hskip 0.12in} c@{\hskip 0.12in} c@{\hskip 0.12in} c@{\hskip 0.12in} c@{\hskip 0.12in}|}
    {\   } & {66.7} & {73.3} & {76.7} & {80.0} \\
    {70.0} & {\   } & {50.0} & {56.7} & {60.0} \\
    {70.0} & {56.7} & {\   } & {63.3} & {66.7} \\
    {76.7} & {76.7} & {66.7} & {\   } & {60.0} \\
    {90.0} & {83.3} & {66.7} & {36.7} & {\   } \\
\end{tabular}
&
\begin{tabular}{|c@{\hskip 0.12in} c@{\hskip 0.12in} c@{\hskip 0.12in} c@{\hskip 0.12in} c@{\hskip 0.12in}|}
    \      & {66.7}  & {86.7}  & {96.7} & {100 } \\
    {73.3} & \              & {56.7}  & {76.7} & {96.7} \\
    {76.7} & {80.0}  & {\   }  & {80.0} & {100 } \\
    {93.3} & {90.0}  & {90.0}  & {\   } & {93.3} \\
    {100 } & {96.7}  & {96.7}  & {86.7} & {    } \\
\end{tabular}
\vspace{2.5 mm}
\end{tabular}
\caption{Confusion matrices for skewness(left), kurtosis (middle) and DCNN (right), for noiseless images (top) and noisy images with moderate noise level $\sigma_{\kappa}^{\rm{space}}$ (middle) and high noise level $\sigma_{\kappa}^{\rm{ground}}$ (bottom). Each entry $\mathrm{CM}_{ij}$ in the confusion matrix is the percentage of maps with true model $i$ classified correctly against model $j$.}
\label{tab:confusion_matrices}
\end{centering}
\end{table}

\paragraph{Mean discrimination efficiency $\rm{MDE}$. } The results for the three variants of the DCNN approach, as well as the other statistics, are summarized in Figure~\ref{fig:mean_discrimination_efficiency}. We plot the $\rm{MDE}$ for skewness and kurtosis calculated from the maps with and without smoothing. In the noise-free case, the DCNN approaches performs the best, although the results are only slightly better than those from skewness and kurtosis statistics. As we start increasing the level of noise, the results of the DCNN become significantly better.
For a level of noise $\sigma=0.42$, all three DCNN approaches outperform the skewness and kurtosis approaches (both smoothed and unsmoothed).
For high levels of noise, the skewness and kurtosis clearly benefit from smoothing the maps. However, their performance is still far behind that of the DCNN.

\paragraph{Pairwise discrimination efficiencies.} The discrimination efficiencies between each pair of models for DCNN and the skewness statistics are presented in Table~\ref{tab:confusion_matrices}. The results for all three variants of DCNN being similar, we only report results for our default method, \emph{DCNN}, which achieved slightly better accuracy at the beginning of the optimization process. The same trends as in Figure~\ref{fig:mean_discrimination_efficiency} can be observed in the confusion matrices. While in the noiseless case the DCNN perform only slightly better than the skewness statistics, the difference becomes very significant for higher levels of noise.
For moderate noise level corresponding to space observations and with a number of galaxies per $\rm{arcmin}^2$ equal to $n_g^{\rm{space}}=100$, the DCNN approach manages to maintain good performance. The accuracy decreases slightly only for  adjacent models (i.e. models whose parameters are closer between each other). The decrease of performance for the skewness and kurtosis statistics is significantly worse for both adjacent and non-adjacent models.
For the highest noise level ($n_g^{\rm{ground}}=10$), the accuracy of all methods drop significantly. The DCNN approach, however, manages to maintain a good performance for non-adjacent models, while the performance of other statistical methods decrease significantly for many pairwise comparisons and becomes close to 50\% (i.e. the same as random guessing).

Finally, we compare the potential ability of the DCNN approach and the 2-pt functions to break the $\Omega_m - \sigma_8$ degeneracy.
We calculated the uncertainty on the $B_8$ from the fiducial tomographic cosmic shear analysis from KiDS using 450 deg$^2$ \citep{hildebrandt2017kids} and obtained $\sigma(B_8)=0.1$.
DES SV tomographic analysis with 139 deg$^2$ \citep{des2015cosmology} gives $\sigma(B_8)=0.2$.
In order to estimate this uncertainty for the DCNN approach, make a simple, order-of-magnitude calculation.
We calculate a factor $p_{ij} = 1-\mathrm{CM}_{ij}/100$, where $\mathrm{CM}_{ij}$ is an entry in the confusion matrix, and set the diagonal $p_{ii}=1$.
This factor $p_{ij}$ will be roughly proportional to the probability density for model $j$, if model $i$ is the true one.
We assume model 2 to be the truth, as its corresponding $B_8$ value is the closest to the one with maximum probability from \citep{hildebrandt2017kids}.
We use the $B_8$ values of the five models, to linearly interpolate and extrapolate their corresponding $p_{ij}$ to the full range of $B_8$.
For this interpolated, approximate probability model, we calculate $\sigma(B_8)=0.1$ from the DCNN approach for the high noise case, and $\sigma(B_8)=0.05$ for the moderate noise.
That approach used only $\sim 4$ deg$^2$ to perform the analysis.
For that reason these uncertainties are not directly comparable; additional differences include the noise level (8.53 galaxies/arcmin$^2$ for KiDS \citep{hildebrandt2017kids}, 5.7 galaxies/arcmin$^2$ for DES \citep{jarvis2016des}, and 10 galaxies/arcmin$^2$ used here), and median redshift of source galaxies (0.53 for KiDS \citep{kuijken2015gravitational}, 0.72 for DES \citep{bonnett2016redshift} and single source plane at redshift $z=1$ used here).
Nevertheless, despite these differences, these results indicate that the DCNN is a promising approach for breaking the $\Omega_m - \sigma_8$ degeneracy.


\section{Network analysis}
\label{sec:network_analysis}

While Deep Neural Networks learn complex nested representations of the data allowing them to achieve impressive performance results, it also limits our understanding of the model\footnote{Recall the term model was described in Section~\ref{sec:deep_network_nomenclature}.} learned by the neural network. In this section, we report various empirical measures that provide some insights into the inner working of this model.

\begin{figure}
\centering\includegraphics[width=0.8\textwidth]{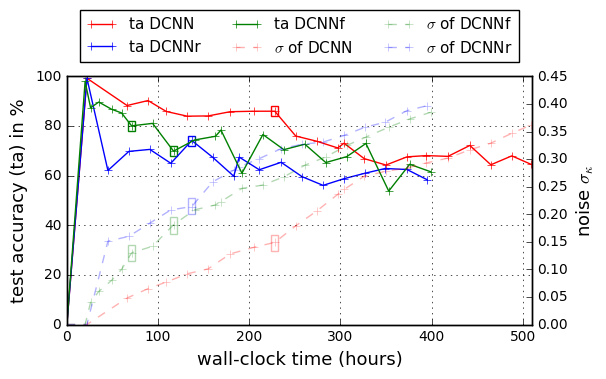}
\caption{Evolution of test accuracy (solid lines) and noise level (transparent dashed lines) over time. Keeping the accuracy at a relatively higher level requires slower increase of noise.\\}
\label{fig:noise_evolution}
\end{figure}

\begin{figure}[h]
\centering\includegraphics[width=1\textwidth]{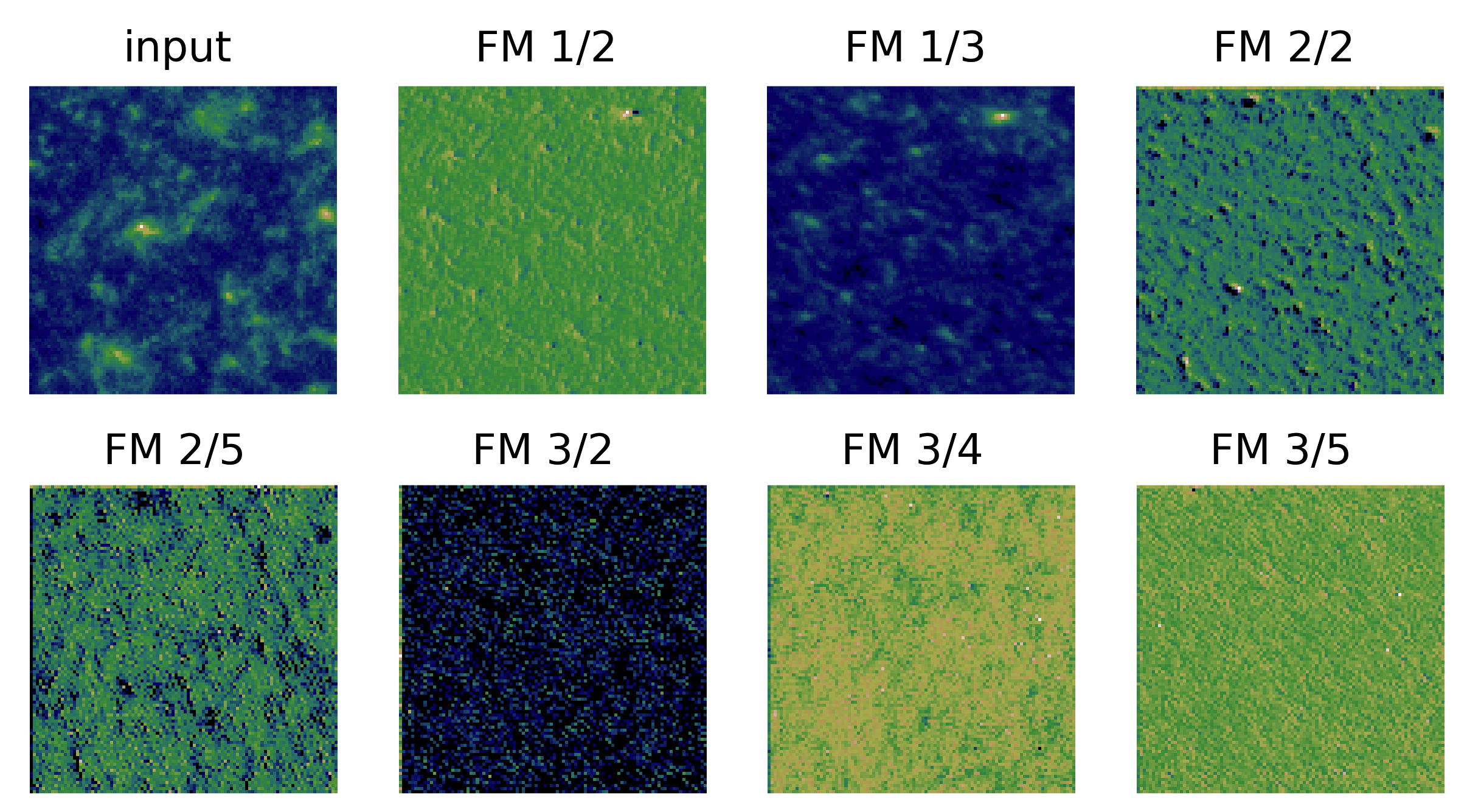}
\caption{Original input (noiseless) and selected feature maps of the network trained on noiseless weak lensing convergence mass maps, each down-sampled to $100 \times 100$ pixels. We see how the network tries to pick out peaks (FM 2/2) or tries to enhance contrast (FM1/3). In deeper layers (e.g. FM 3/5), the feature maps are not easily interpretable.}
\label{fig:fm_clean}
\end{figure}
\begin{figure}
\centering\includegraphics[width=1\textwidth]{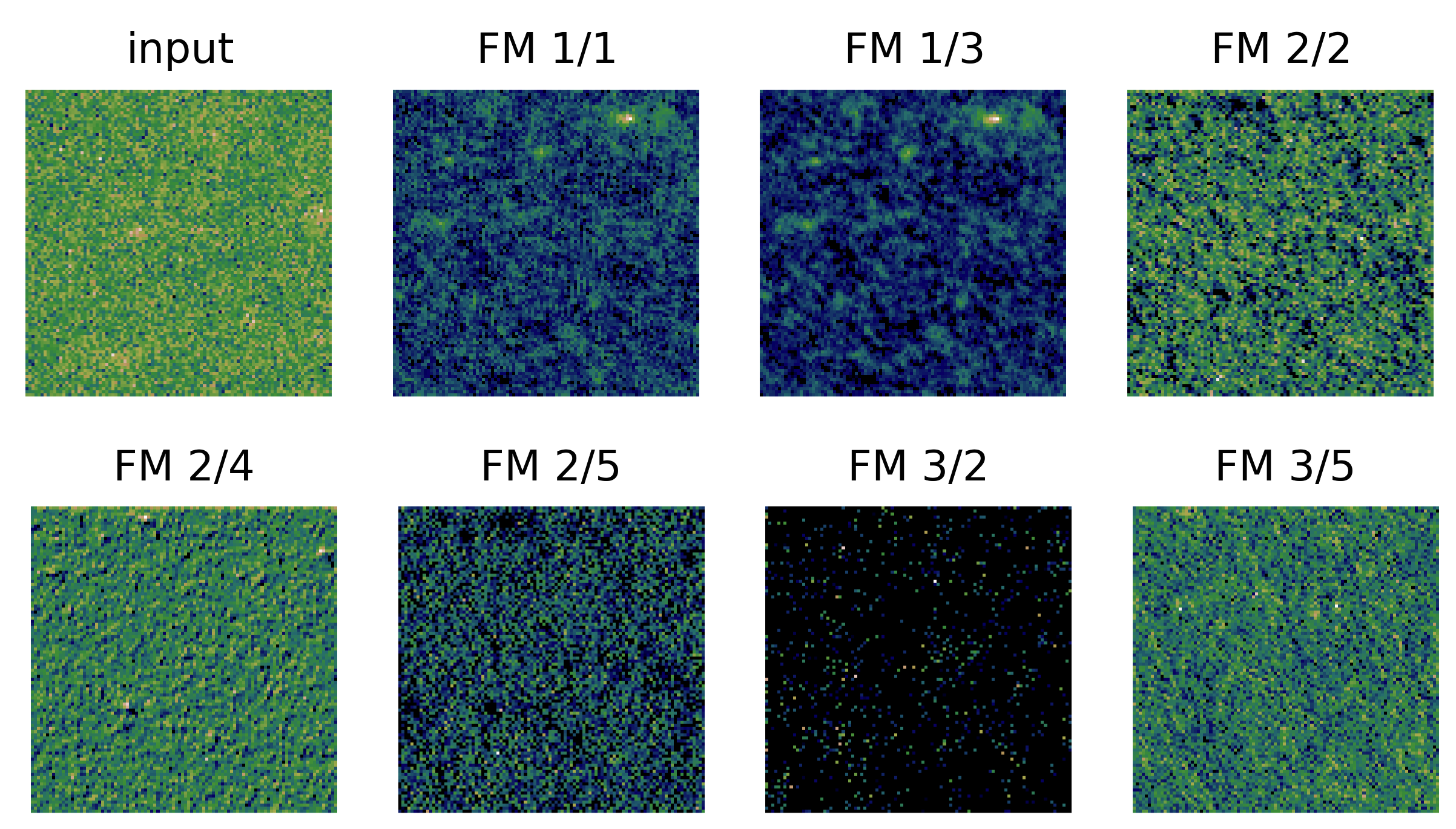}
\caption{An example original input convergence map ($\sigma _{\kappa} =0.13$) and selected feature maps of the trained network, each down-sampled as above. The feature maps are not expected to match the example convergence map, but could have similar texture.
The feature maps after the first convolution (FM 1/1 and FM 1/3) show a similarity to noisy maps, convolved with a Gaussian kernel (de-noising). After the second convolution (FM 2/2 and FM 2/4), the network identifies peaks again, before in later layers interpretation becomes more difficult (e.g. FM 3/2).}
\label{fig:fm_noisy}
\end{figure}

\subsection{Learning curve and redundancy of accuracies}
\label{sec:learning_curve}

We first report the accuracy of the network during the course of training in Figure~\ref{fig:noise_evolution}. We observe that the fast increase in the noise level for DCNNf results in a worse performance at the beginning. However, the results after 400 hours of training seem to suggest, that the three approaches, despite their differences, converge to similar performance. We also experimented with a non-adaptive strategy where a constant amount of noise is added to the input. This experiment didn't show any progress for approximately 40 000 iterations during which the accuracy was fluctuating at $\sim0.2$, thus justifying the choice of the noise adaptive strategies described in this manuscript.

Upon closer investigation, we also notice that a significant drop in training accuracy occurs after the noise level is increased. After more training, the network's performance increases again until the next increase in noise.

\subsection{Visualisation of learned filters and feature maps}

We can also consider the filters and the feature maps learned in several convolutional layers of the network as done in \citep{Lecun2015}. These are shown in Figure~\ref{fig:fm_clean} where FM x/y is the feature map of the y-th filter in the x-th hidden layer.
The upper left panel contains an example convergence map. The feature maps are not expected to look similar to the example map; in particular the clumps are not expected to overlap.
While some of the feature maps (e.g. FM 1/2, FM 2/2) seem to have a similar texture to the peaks in the input data, others seem to enhance the contrast (e.g. FM 1/3). In deeper layers, it is harder to interpret what kind of representations the network is trying to find (e.g. FM 3/2). Similar observations were made for the network trained on noisy data (Figure~\ref{fig:fm_noisy}): FM 1/1 as well as FM 1/3 appear like de-noised convergence maps (similar to enhanced contrast on discriminating features), while FM 2/2 and FM 2/4 seem to resemble peaks in the data. The internal representations learned in deeper layer is again too complex to be able to be easily interpreted.


\section{Conclusion}
\label{sec:summary}

We proposed a novel DCNN approach to discriminate cosmological models using noisy convergence mass maps.
The algorithm was designed to capture non-Gaussian information from the mass maps, and thus breaking the degeneracy between $\Omega_m$ and $\sigma_8$ parameters, which is characteristic for Gaussian statistics, such as power spectrum or correlation functions.
The network is trained on a set of realistic simulations, all of which were designed to follow the degeneracy in the $\Omega_m$ - $\sigma_8$ plane, and thus having similar power spectra. Training a DCCN was observed to be challenging due to the high level of noise added to the convergence maps to simulate realistic observatory settings. We therefore designed a novel training strategy which helps the network to handle the high level of noise. This is achieved by starting with noise-free data and gradually increasing noise during the course of training. We also conducted a through examination of various adaptation of this strategy showing how they impact the performance of the network.

We compared the performance of DCNN to several commonly used higher order statistics, namely skewness and kurtosis. While the performance of DCNN is close to the skewness statistics in the noise-free setting, the situation is different in the setting with high noise levels: the DCNNs can clearly cope with the noise and significantly outperform other statistics.
In the high noise setting, the mean discrimination efficiency was greater than $85\%$ for the DCNN approach, while other methods achieve less than $70\%$ efficiency.
This result indicates that Deep Learning approaches can potentially be a powerful statistic for constraining cosmological models, as it is able to extract non-Gaussian information from the data.
It is indeed able to break the degeneracy between $\Omega_m$ and $\sigma_8$ from weak lensing, and constrain the parameter space in the $B_8$ direction.
Finally, we estimate that the DCNN approach can give significantly improved constraints on the $B_8$ parameter than the two point statistics.

Our results demonstrate the high potential of neural networks for the analysis of real data from cosmological surveys.
There are various practical aspects that have to be considered to further extend this algorithm to a more realistic case.
We have shown that the DCNN approach can discriminate between a relatively small discrete set of cosmological models, and the natural next step is to develop a CNN which will be able to
map to a continuous range of cosmological parameters.
Other considerations will surely include the sensitivity of the method to discrepancies between the real observed data and the simulations.
These discrepancies can be caused, for example, by insufficiently well modelled observational effects.
An Neural Network could be designed, such that it is invariant to the uncertainty on these effects.

One significant advantage of this approach is its ability to construct features directly from data, thus alleviating the need to create hand-designed features. The visualization of the filters and the feature maps proves the DCNN approach captures complex salient statistics from the data.


\acknowledgments

This work was supported by a grant from the Swiss National Supercomputing Centre (CSCS) under project ID \emph{da02 - Cosmology \& Deep Learning} for the machine learning part as well as by a further grant from the Swiss National Supercomputing Centre (CSCS) under project ID \emph{p501 - PASC Cluster - Monch} for the simulations and data acquistion.
This work was also supported in part by grant number $\rm{200021\_169130}$ from the Swiss National Science Foundation.
 \\

\bibliography{refs}{}
\bibliographystyle{unsrt}

\appendix
\appendix
\section{N-body simulations and convergence maps calculation}
\label{sec:convergence_map_calculation}

For scalability reasons, we used a fast simulation code, \textsc{L-PICOLA} \citep{Howlett2015a}, which generates and evolves a set of initial conditions into a dark matter field much faster than a full non-linear N-Body simulations. \textsc{L-PICOLA} starts time stepping from a pre-set initial redshift $z_{\rm{init}}$, which we set to $z_{\rm{init}} = 9.0$ as recommended in \citep{Howlett2015a} and simulates the gravitational interaction and evolution of the contained particles over time.
Our simulations then return snapshots of particle positions and velocities in a box of length $L$ (see Table \ref{tab:5cosmparams}) at pre-set redshifts. To ensure an efficient calculations of the convergence maps, 15 redshifts have been chosen, such that $15 \times L = C\left(0,1\right)$, where $C\left(z_1 ,z_2 \right)$ defines the comoving distance between redshift $z_1$ and $z_2$ respectively. The comoving distance is a time independent distance measure in astrophysics, that factors out the influence of the expansion of the universe. Since we want to produce convergence maps with a field size of $\fieldsize$
(and $3.81 \approx \arctan \left( 1/15 \right)$), the cosmological parameters also lead to the specific box length of each model:
$L_{\rm{model}}=C_{\rm{model}} ( 0, z_0 ) / 15 \ \cdot \ h_{\rm{model}}$.

Weak lensing convergence mass maps were obtained by integrating density fluctuation along the line of sight using the Born approximation, which approximates the propagation of light rays as straight lines. The simulations output boxes of predefined length $L$ with particle positions and velocities at given redshift $z_i$. These boxes can be projected along one axis onto a 2D plane. Calculating the weighted sum of these projections at different redshifts according to equation \ref{eqn:cmapeqn} defines the effective per pixel convergence $\kappa _e$, calculated as

\begin{align}
    \kappa _e \approx \frac{3 H_0 ^2 \Omega _m L}{2c^2} \sum _i \frac{\chi _i\kll\chi _0 - \chi _i \klr}{\chi _0 a\kll \chi _i\klr} \kll \frac{n_p R^2}{N_t s^2} - \Delta r_{f_i}\klr
    \label{eqn:cmapeqn}
\end{align}
where $c$ is the speed of light, $\chi_i$  is a comoving distance from source to the observer, $\chi _0$ is the comoving distance to the source galaxies. The source galaxies are placed on a single source plane at redshift $z=1$. Technical parameters dependant on the simulation are: the number of particles contained within one pixel $n_p$, resolution of the maps in $R \times R$ pixels, $N_t$ the total number of particles contained in the simulation and $\Delta r_{f_i} = (r_2 -r_1)/L$ with boxlength $L$ and comoving distances $r_{1,2}$. The ratio of the length of the lensing plane $L_p$ and the boxlength is $s=L_p/L$.

By exploiting the simulated data explained in Section~\ref{sec:simulations} and inserting it into the Equation~\ref{eqn:cmapeqn}, we get the final weak lensing convergence mass maps, which we used for our further work.
Examples of typical maps for each of the five different cosmological models can be seen in Figure~\ref{fig:example_maps}.

In a mass maps reconstructed from the observed data, we expect to find noisy mass maps due to measurement errors and intrinsic ellipticity dispersion of galaxies. In order to obtain noisy weak lensing mass maps, we added random Gaussian noise to the convergence. Following the example of \citep{pires2009model}, the noise levels of convergence $\kappa$ were determined by $ \sigma ^2 _{\kappa} = \sigma ^2 _{\epsilon} / (A n_g)$, where $\sigma _{\epsilon}$ is the root mean square of the shear distribution and assumed to be $\sigma _{\epsilon} \approx 0.3$, $n_g$ is the average number of galaxies per arcmin$^2$ and $A$ is the pixel size in arcmin$^2$. Two noise levels were of special interest: a very optimistic space observation case, where we assumed $n_{g}^{space} = 100 \ \rm{galaxies} / \rm{arcmin}^2$ and a realistic ground observation case with $n_{g}^{\rm{ground}} = 10 \ \rm{galaxies}/\rm{arcmin}^2$. These correspond to $\sigma_{\kappa}^{\rm{ground}}=0.42$ and $\sigma_{\kappa}^{\rm{space}}=0.14$

For each of the cosmological simulations, we generated 2500 convergence maps. This has been done by following procedure: at first, we chose snapshots randomly from one of our simulations (1 out of 15) for the needed redshift. Then we chose a random orientation of the axis (1 out of 6 combinations, without flips) and finally chose a random area from each box. Neglecting the randomization of the
clipping of snapshots, this yields $6 \times \sum_{i=1}^{i=15} i = 990$ different possibilities to combine draws of snapshots from the different simulations. This yields 2500 maps, which can be considered close to being statistically indepdendent.

\end{document}